\documentclass[preprint,aip,showpacs]{revtex4-1}

\usepackage{graphicx, dcolumn, bm, epstopdf}
\makeatletter

\newcommand{\Rmnum}[1]{\expandafter\@slowromancap\romannumeral #1@}
\usepackage[breaklinks=true,colorlinks,citecolor=blue,linkcolor=blue,urlcolor=blue]{hyperref}
\setcitestyle{super}
\makeatother

\begin{document}

\title{Anomalous ferroelectric switching dynamics in single crystalline SrTiO$_3$}

\author{Vinay Kumar Shukla}
\email{vkshukla@iitk.ac.in}

\author{Soumik Mukhopadhyay}
\email{soumikm@iitk.ac.in}
\affiliation{Department of Physics, Indian Institute of Technology, Kanpur 208 016, India}


\begin{abstract}
Pure SrTiO$_3$ in bulk form is known to be an `incipient ferroelectric' where quantum fluctuations of lattice positions prevent long range ferroelectric ordering at finite temperature. We show evidence and identify the origin of ferroelectric relaxation up to nearly room temperature in single crystalline $SrTiO_3$. Strikingly, the origin of the observed ferroelectric switching is intrinsic that is coherent switching of surface nanopolar regions and not due to the nucleation and growth of domains, as described by Kolmogorov-Avrami-Ishibashi (KAI) Model.
\end{abstract}


\maketitle

\section{Introduction}
Apart from its traditional usage as substrate for epitaxial growth of complex oxides, Strontium Titanate ($SrTiO_3$) is widely regarded as a promising gate dielectric to replace $SiO_2$ in nano-electronic applications due to its
high dielectric constant at room temperature and possibility of its integration with the Silicon substrate~\cite{McKee, Forst}. Interestingly, $SrTiO_3$ is also rapidly gaining importance as substrate for transport study in Graphene~\cite{Raymond}, as a preferred template for creating 2D electron gas (2DEG) and superconductivity at oxide interfaces~\cite{Thiel, Reyren}, as exhibiting properties of a highly metallic 2DEG at its vacuum-cleaved surface and as a potential ferroelectric~\cite{Jang}. The present article discusses the nature and origin of ferroelectric switching in single crystalline $SrTiO_3$ (STO).

Almost $25$ years back, Bickel \textit{et al.} provided evidence of ferroelectric relaxation due to STO surface reconstruction, the so called `puckering' of oxygen ions, using low energy electron diffraction~\cite{Bickel}. Since the `incipient' ferroelectric (FE) phase in STO is subjected to competing interactions such as quantum fluctuations and anti-ferro distortions~\cite{Zhong, Hiromi}, any small external perturbations such as cation doping, isotope substitution, stoichiometry deviation etc. would affect its FE properties. M.Itoh \textit{et al.}~\cite{Itoh, Taniguchi} observed that low temperature (below 23 K) ferroelectricity was induced in single crystalline STO by isotope exchange of $O_{16}$ by $O_{18}$. Low temperature ferroelectricity can be induced in STO by cation doping (such as Ca) and by mechanical stress as reported by Bednorz \textit{et al.}~\cite{Bianchi1,Bianchi2} and others~\cite{Mitsui, Burke}. The appearance of ferroelectric instability in STO has also been reported near antiphase domain boundaries at low temperatures below the antiferrodistortive transition~\cite{Alexander}. The flexoelectric effect i.e. the coupling between strain gradient and dielectric polarization could also lead to ferroelectricity in STO~\cite{Zubko, Anna}. The ferroelectricity in STO can also be induced by applying external electric field. Fleury \textit{et al.}~\cite{Fleury1, Fleury2} have observed first order raman modes in single crystalline STO, albeit at low temperature, in presence of external electric field which removes the center of inversion symmetry of the crystals.

In recent years quite a few reports have been published regarding the existence of ferroelectricity in $SrTiO_3$ (STO), which has renewed the debate on the nature of `ferroelectricity' in $SrTiO_3$ in various forms and under different experimental conditions. The strong and asymmetric hysteretic response in the gate voltage dependence of resistivity in Graphene placed on STO are attributed to the ferroelectric dipole moment arising out of the `puckered' surface~\cite{Raymond}. In 2004, room temperature ferroelectricity was reported in epitaxial thin films of STO, characterized by a pronounced peak in the dielectric constant measured in plane at microwave frequencies~\cite{Haeni} and the relaxor nature of ferroelectricity was established later on by the same group~\cite{Biegalski}. Bi-axial strain plays a key role in inducing ferroelectricity in thin films of STO. Biegalski \textit{et al.}~\cite{Biegalski} reported  strain induced relaxor ferroelectricity in STO thin films grown on $DyScO_3$ substrate, which was likely due to Sc doping. Furthermore, by using the direct comparison of the strained and strain-free STO thin films grown on STO and $NdGaO_3$ substrates, Jang \textit{et al.}~\cite{Jang} reported that all $SrTiO_3$, including bulk single crystal systems, were relaxor ferroelectrics arising out of minute non-stoichiometry in nominally stoichiometric STO and the role of strain was merely to stabilize the long range correlation of pre-existing nanopolar regions. The equivalent role of biaxial strain and electric field in stabilizing the polar domains were also discussed. The Sr/Ti stoichiometry played a key role was also pointed out by Tenne \textit{et al.}~\cite{Tenne} who reported appearance of first order polar Raman modes in non-stoichiometric STO thin films. Herger \textit{et al.}~\cite{Herger} used surface x-ray diffraction to show that $TiO_2$-terminated STO could exhibit surface ferroelectricity.

In general, there is an agreement that the observed FE like properties of STO thin films or single crystals are relaxor-type with presence of polar nanoclusters~\cite{Haeni, Warusa}. Whether these nanoclusters can give rise to net macroscopic electric polarization is still an open question. Very recently, however, Lee \textit{et al.} have come out with evidence that alignment and stabilization of these polar nanoregions in ultra-thin $SrTiO_3$ film is indeed possible under electric field which can result in net ferroelectric polarization at room temperature~\cite{Lee} although the dynamics of the switching process is not elaborated. Interestingly, majority of the experiments on $SrTiO_3$ have been carried out without using any electrode material which does not help establish a holistic insight into the subject since the electrical boundary conditions at the interface influence the ferroelectric phase. In fact the effective screening length at the metal-oxide interface might be crucial in stabilizing ferroelectricity~\cite{Spaldin, Stengel1, Stengel2}. In this article, we demonstrate that one can observe net switchable polarization even in single crystalline $SrTiO_3$ with metal electrodes which arises out of coherent switching of nanopolar domains under external electric field.

\section{Experimental Details}
The Dielectric and ferroelectric properties of $SrTiO_3 (100)$ single crystal substrate were studied using Radiant Ferroelectric Tester (Precession Premium II), which uses virtual ground circuitry, in the in-plane geometry with silver paint, evaporated Silver and Gold as electrodes. The Sr/Ti ratio in the STO (100) substrate was characterized using Energy Dispersive X ray Spectroscopy (EDS) and we found evidence of Sr deficient and Sr enriched regions (with maximum deviation of around $1.8\%$) over few microns length scale which averages out to be stoichiometric over larger length scale of few $100$ microns, still well below the separation between the electrodes. Atomically smooth surfaces were observed on STO (100) using Atomic Force Microscopy (AFM) (not shown).

The switching polarization vs. electric field loops and the associated dynamics in real time were measured using a set of $5$ `monopolar' triangular voltage pulses, the underlying principle being similar to the standard ``PUND, Positive Up Negative Down'' method~\cite{Naganuma} alternatively known as ``Double Wave method''~\cite{Fukunaga}. The first pulse is treated as preset pulse to align the polarization of the sample in a particular direction. The second triangular pulse of given amplitude $V_0$ and timescale $t_0$ is applied in a direction opposite to that of the preset pulse. This gives the half hysteresis loop associated with electric polarization which includes both switching and non-switching components. In the third step, a `non-switched' pulse of same amplitude and duration is applied in the same direction as that of the preceding pulse, which leads to half hysteresis loop containing only the non-switched polarization. The difference between half hysteresis loops measured in sequence 2 and 3 retains the switching polarization and gets rid of the non-switching part. Similarly, sequence 4 and 5 are repeated but in opposite direction. Combining the difference between half hysteresis loops measured in sequence 2 and 3 on one hand and 4 and 5 on the other gives us the complete four-quadrant $P_{SW}$ vs. $E$ hysteresis loop. The capacitance was measured as described in Ref~\cite{vkshukla}. The temperature dependent polarized Raman Spectroscopy was carried out using STR Raman Spectrometer (Airix Corporation, Japan).

\section{Results and Discussion} 
\subsection{Dielectric Spectroscopy}

\begin{figure}
  \includegraphics[width=9cm]{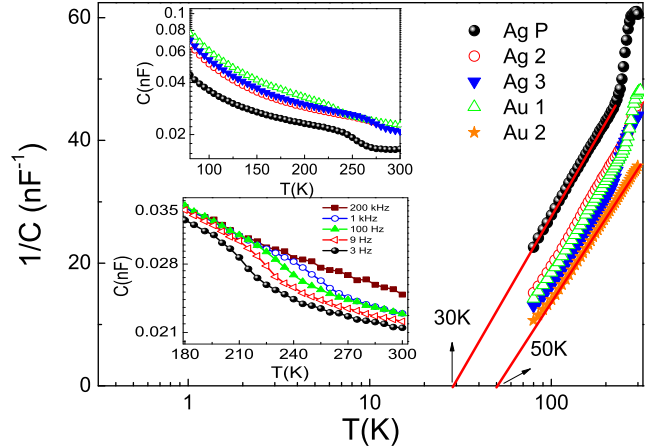}\\
  \caption{The inverse of capacitance vs. temperature curve for different contact electrodes. Inset (above): The temperature dependence of capacitance for different contact electrodes at f=100 Hz. Inset (below): The temperature dependence of capacitance for the sample with Ag paste electrodes at different frequencies.}\label{fig:dielec}
\end{figure}
We present the temperature dependence of capacitance of single crystalline $SrTiO_3(100)$ for different types of contacts: first, $SrTiO_3$ with 1mm separated silver paint electrodes (denoted as Ag P); second, silver electrodes deposited by e-beam evaporation with probe separation of 500 micron and 50 micron denoted as Ag 2 and Ag 3 respectively; third, e-beam deposited gold electrodes namely Au 1 and Au 2 for 500 micron and 50 micron probe separation. The temperature dependence of capacitance is not influenced by the choice of electrodes. Strikingly, all the samples display dielectric anomaly near room temperature and strong dispersion at low frequencies around the anomaly (Fig.~\ref{fig:dielec}). To our knowledge, such dielectric anomaly near room temperature has not been reported before. The switching polarization is strongly dependent on the voltage pulse width and temperature, which saturates in the high pulse width regime. The I-V characteristics at all temperature and for all electrodes were linear and symmetric suggesting that the contact is ohmic, thus ruling out artifact arising out of Schottky junction formation. Measurements were done both in the out-of-plane and in-plane contact geometry. However, the observed dielectric anomalies are not affected by the geometry of the contacts. Semi-log plot of capacitance versus temperature (Fig.\ref{fig:dielec}) suggests significant deviation from the conventional Curie-Weiss behavior. The calculated value of relative permittivity of $SrTiO_3(100)$ is 1120 at 100 K, consistent with previous report~\cite{Neville}.

In general, dielectric anomalies are usually attributed to structural phase transitions. However, this is not the case here because the $SrTiO_3$ undergoes anti-ferrodistortive transition from cubic to tetragonal phase at 105K and classical to quantum paraelectric phase at 37K and no such structural transition is associated with the high temperature regime~\cite{Cowley}. Moreover, the broad dielectric relaxation observed around 225 K as shown for Ag 3 in lower inset of Fig.~\ref{fig:dielec}, shifts towards higher temperature with increasing frequency eventually disappearing altogether. Such behaviour is usually attributed to the relaxor ferroelectrics. However, the temperature dependence of capacitance neither follows Curie-Weiss behaviour nor exponential trend as discussed by Cheng \textit{et al.}~\cite{Cheng}. The origin of this frequency dependent dielectric anomaly could be attributed to the Maxwell-Wagner relaxation which arises due to trapping of charges at the sample-electrode interface~\cite{Jianjun} at low frequencies while vanishing at higher frequencies. Interestingly, the capacitance values for all contacts are nearly the same. To investigate the ferroelectric response of single crystalline $SrTiO_3$ with different contact electrodes and to explore the possibility of any dipolar relaxation associated with the observed dielectric anomaly, we have carried out remanent electric polarization measurements.

\subsection{Remanent electric Polarization measurements}

\begin{figure}
  \includegraphics[width=8.7 cm]{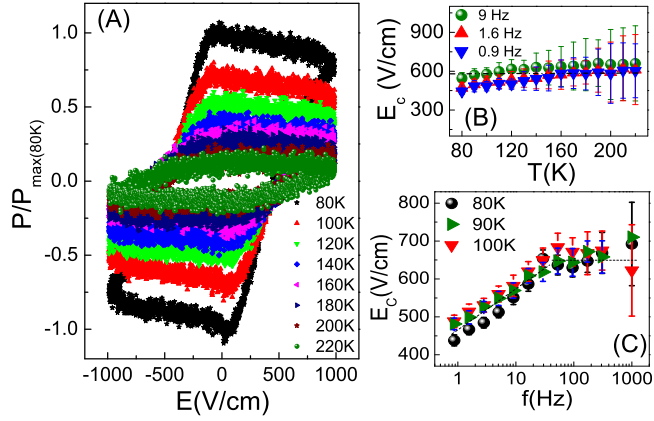}\\
  \caption{(A)Evolution of the ferroelectric P-E loop with temperature in the in-plane geometry for AgP. The switching polarization decreases with temperature but the coercivity remains unchanged,(B)The frequency dependence of the switching field showing power law behavior,(C) The lack of any variation in switching field with temperature is shown. }\label{fig:hyst}
\end{figure}

Traditional P-E loops are generally composed of ferroelectric, parasitic, or stray capacitance and conductive contributions. Therefore, PUND method is a better tool to explore the remanent electric polarization as discussed in our previous report~\cite{vkshukla}. Instead of rectangular pulses applied in capacitive measurements, here, we have applied triangular pulses so that rate of change of electric field over time should be fairly constant except that for reversal points.
Different pulse widths ranging from 1 Hz-1 kHz were applied. Fig.~\ref{fig:hyst}A shows temperature variation of remanent polarization normalized with respect to it maximum value at 80 K ($P_{max(80 K)}$)  versus electric field loops for Ag P sample at an applied field 1 kV/cm having pulse width of 1.5 Hz. The observed coercive field is close to 400 V/cm at 80 K. To see the effect of varying pulse width on the saturation polarization and shape of the PE loops, we have taken loops by varying pulse width at different temperatures as shown in Fig.~\ref{fig:hyst}A. With increasing time period, the saturation polarization first increases sharply and then saturates above 600 ms (not shown), while on the other hand, the same decreases with increasing temperature, typical of conventional ferroelectric materials. The observed ferroelectricity is not of flexoelectric origin as the percentage change in polarization is much higher, nearly 90 \% as compared to 60 \% change in capacitance in the same temperature range. Moreover, there is no externally applied strain gradient (typically strain gradients of magnitude $\sim$ $10^{-6}$ is employed to induce flexoelectricity~\cite{Zubko}). The typical value of $P_{max(80 K)}$  is 6.5 $\pm$ 2.5 $\mu$C/$cm^2$ at pulse width of 1 kHz, whereas it reaches to value of 25.5 $\pm$ 5.5 $\mu$C/$cm^2$ at pulse width of 1 Hz. We have observed marginal decrease in saturation polarization with increasing electric field (Fig.~\ref{fig:hyst}A)  at low frequencies, which is also reported by others groups~\cite{HoWon,Hongyang}.  
	
We have further investigated the value of coercive electric field ($E_c$) which is extracted from the observed remanent polarization versus electric field loops. In order to avoid the possibility of artifacts, $E_c$ should be less than the maximum applied electric field. In general, coercivity should decrease with increasing the temperature as expected within the framework of thermally activated nucleation and growth of domains in presence of external electric field. Contrary to the expectation, the temperature dependence of coercive field (as shown for Ag P in Fig.~\ref{fig:hyst}B) indicates that coercivity does not decrease with increasing temperature. Fig.~\ref{fig:hyst}C shows the frequency dependence of coercivity for Ag P which obeys a typical power law behaviour $E_c \sim f^\alpha$ below 40 Hz. Observed value of ${\alpha = 0.11}$ is in close agreement to that obtained by J. F. Scott~\cite{Scott} for conventional ferroelectric films. Above 40 Hz, remanent polarization decreases with increasing pulse frequency making it difficult to estimate the coercive field values precisely. 

\subsection{Switching kinetics of remanent electric polarization}

The traditional approach to explain the FE switching kinetics under an external field $E_{ext}$, often called the Kolmogorov-Avrami-Ishibashi (KAI) model~\cite{Ishibashi}, is based on the classical statistical theory of nucleation and unrestricted domain growth. For a uniformly polarized FE sample under $E_{ext}$, the KAI model gives the time (t)
dependent change in polarization $\Delta P(t)$ as
\begin{equation}\label{eq:1}
\Delta P(t)= P_{s}[1-exp\{-(t/\tau_0)^n\}]
\end{equation}
where $n$ and $\tau_0$ are the effective dimension and characteristic switching time for the domain growth, respectively,
and $P_s$ is spontaneous polarization. 

\begin{figure}
  \includegraphics[width=8 cm]{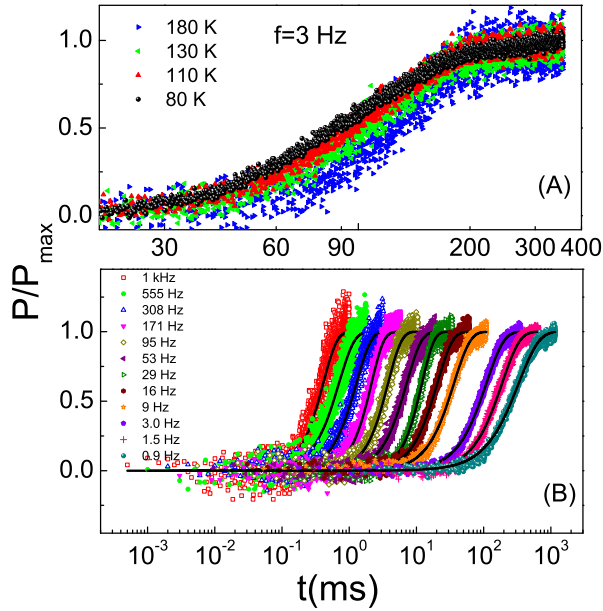}\\
  \caption{(A) The ferroelectric switching for a given frequency $3$ Hz at different temperature showing that the timescale of switching increases with temperature.(B) Ferroelectric switching at $80$ K is shown for a wide range of frequencies. Continuous lines are the fits to the experimental data.}\label{fig:swtemp}
\end{figure}

In general, $\tau_0$ is proportional to average distance between the nuclei,
divided by domain wall speed. The second model is known as the nucleation-limited-switching (NLS) model~\cite{Tagantsev, Jo}, mainly applicable for polycrystalline systems. This model assumes the existence of multiple nucleation centers having independent switching kinetics. The switching in an area is considered to be triggered by an act of the reverse domain nucleation and the switching kinetics is described in terms of the distribution function of the nucleation probabilities in these areas. Here, the coercive field values are within the range $\sim 400-900$ V/cm, much less than that predicted by Landau mean field theory~\cite{Landau}, presumably indicating a switching mechanism which is inhomogeneous and nucleation mediated~\cite{Yang}. We have fitted the time dependent switching polarization curves in Fig.~\ref{fig:swtemp} by KAI model as we do not expect the switching kinetics to be nucleation limited (consisting of independent nucleation centers) for high quality single crystalline samples.

\begin{figure}
\includegraphics[width=8.3 cm]{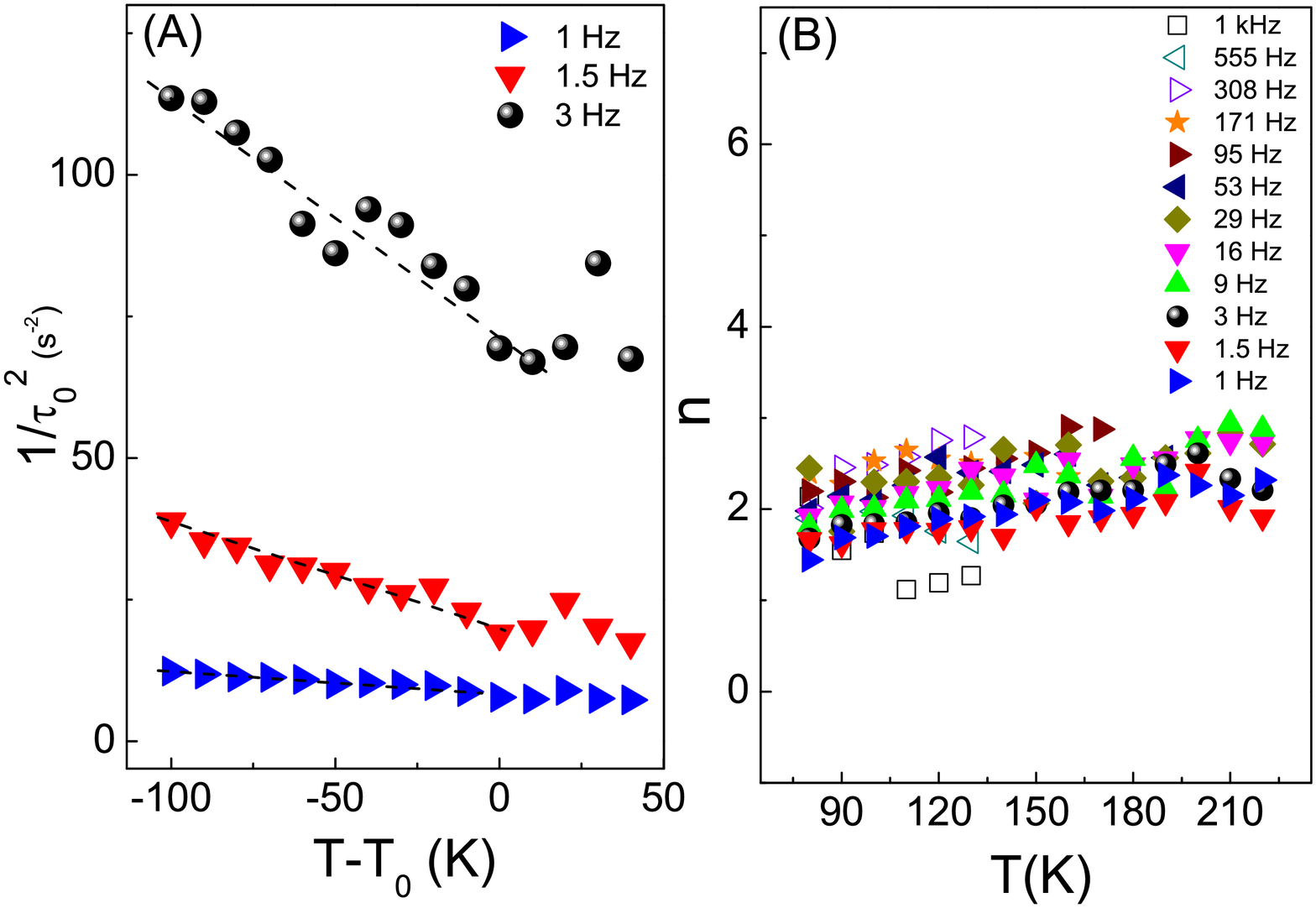}\\
\caption{(A) Characteristic switching time ($\tau_0$) and (B) dimensionality (n) extracted from fitting of Remanent electric polarization versus time by KAI model plotted against temperature for pulse frequencies 0.9 Hz-1 kHz. Maximum applied field is 1 kV/cm.}\label{fig:dimension}
\end{figure}
We have explored the switching kinetics in fairly broad range of frequencies from 1 Hz to 1 kHz. Below 1 Hz, the P-E loops gets distorted having coercivity close to that of maximum applied field and enhanced resistive contributions come into play. Above 1 kHz, the values of remanent polarization decreases significantly making it difficult to measure its value accurately. The intrinsic parameters $\tau_0$ and n obtained for Ag P is shown in Fig.~\ref{fig:dimension}. The characteristic time $\tau_0$ obtained for different applied pulse widths shows gradual increase with temperature up to a certain transition temperature $T_0$ and slight decrease beyond the transition, which is clearly not consistent with KAI model. This indicates that observed ferroelectric polarization may not be coming from nucleation and growth of domains as expected. This is surprising since the interface could easily act as nucleation site because of its high free energy due to the broken symmetry, strain, electric fields, charge, and altered chemical structure, which decrease the nucleation barrier. On the other hand, values of n (dimensionality) obtained from fitting the normalized polarization versus time by KAI $\beta$ model (Equation.~\ref{eq:1})~\cite{Ishibashi} is close to 2. In spite of the lower coercive field values, the temperature dependence of $\tau_0$ (Fig.~\ref{fig:dimension}(A)) follows the behavior of intrinsic switching time described within the framework of Landau-Khalatnikov (L-K) theory in which electric dipoles of the crystal are highly correlated and undergoes coherent switching~\cite{Vizdrik, Ducharme} and is completely different from thermally activated nucleation and growth of ferroelectric domains. The intrinsic switching time in L-K theory can be expressed as a function of coercivity and temperature as,
\begin{equation}
\frac{1}{\tau_0} \sim \left(\frac{E}{{E_C}-1}\right)^{1/2} \left(1-\frac{T-T_{0}}{T_{1}-T_{0}}\right)^{1/2}
\end{equation}
where $T_1$ is defined as $T_1=T_0 + 3\beta^2/4\gamma \alpha$ and $\alpha$, $\beta$, $\gamma$ are the Landau coefficients. Near the transition temperature $T_0$, $E_C$ is almost independent of temperature and $\tau_0$ follows the square root dependence with temperature as shown in Fig.~\ref{fig:dimension}A. Moreover, the order of intrinsic switching time $\tau_0$ is consistent with the earlier experiments on coherent switching~\cite{Ducharme}.

\begin{figure}
  \includegraphics[width=7.2 cm]{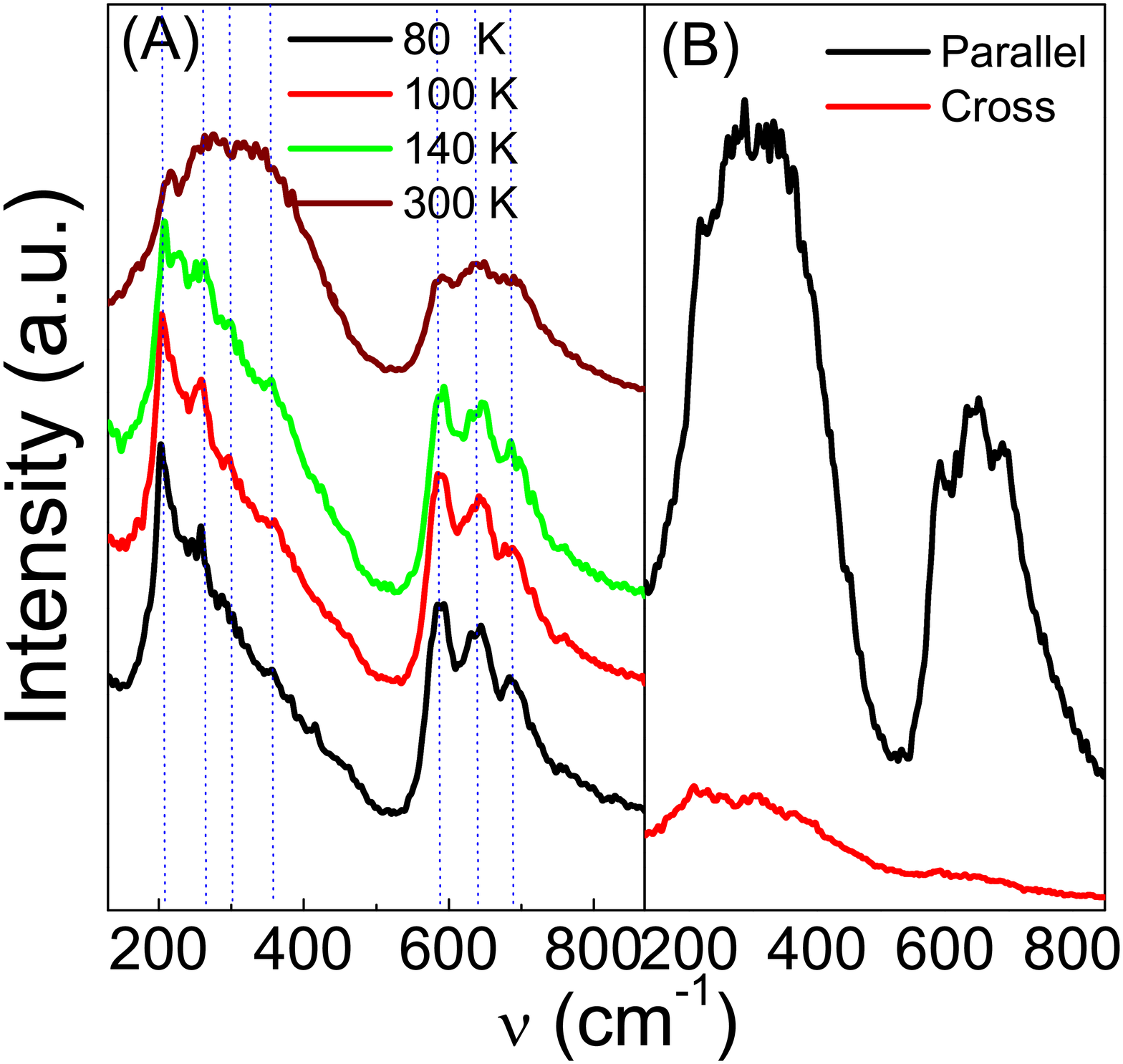}\\
  \caption{(A) Variable temperature Raman Spectra for $SrTiO_3(100)$ from 80 K-300 K. Dotted lines represents the second order Raman modes. (B) Polarized Raman Spectra at 300 K with polarizer and analyzer in parallel and cross positions. }\label{fig:raman}
\end{figure}

\subsection{Raman Spectroscopy measurements}
We also carried out polarized Visible-Raman spectroscopy with DPSS green laser of wavelength 532nm (2.33 eV) for excitation. The laser power density at the sample was 1.25 mW/$cm^2$. A 50 x microscope objective was used with laser spot size of 1 mm. To improve the signal to noise ratio, each spectrum was obtained by the accumulation of ten independent scans with laser exposure time of 10 s per scan having spatial resolution of 2.9 cm$^{-1}$. Fig.~\ref{fig:raman}A shows the Raman spectra of $SrTiO_3$ at different temperatures from 80 K to 300 K. The major spectra consists of second order modes shown by dotted lines with appearance of few less intense peaks below 100 K. These small peaks below 100 K are suggestive of anti-ferrodistortive structural transition of $SrTiO_3$. The appearance of second order Raman modes which arises due to creation or destruction of two phonons is consistent with previous reports~\cite{Nilsen, Sirenko, Jang}. The polarized raman spectra has been taken (shown in Fig.~\ref{fig:raman}B) in parallel and crossed configurations. We have observed significant suppression of Raman signals in the crossed configuration suggesting absence of large scale ferroelectric polarization at room temperature. This is consistent with the EDS result which does not provide any evidence of Sr deficiency over micron size length scales. However, we cannot rule out the existence of nanoscale randomly oriented ferroelectric domains.

It is generally understood that imperfect compensation of the polarization charges at the metal-ferroelectric interface due to the finite electronic screening length of metallic electrodes leads to a depolarizing field that tend to suppress ferroelectricity. However, Stengel \textit{et al.}~\cite{Stengel2} demonstrated, using first principles calculations, that the influence of the local chemical environment at the interfaces of normal metal and perovskites oxides cannot be neglected. The overall effect of including both purely electronic screening and interatomic force constants leads to a covalent bonding mechanism that yields a ferroelectric behaviour of the interface rather than a suppression of the same as predicted by semiclassical theories~\cite{Stengel2}. The prediction that perovskite oxides can show a strong
interfacial enhancement of the ferroelectric properties when weakly bonded to a simple metal should be relevant to the experimental results presented here, especially since $SrTiO_3$ is susceptible to ferroelectric instability at small perturbation. But whether such a scenario should manifest itself in the switching mechanism which involves nucleation and growth of favorable domains or the coherent rotation of the polar domains remains unclear. On the other hand, a recent experiment by Lee et al.~\cite{Lee} used naturally existing polar nanoregions formed due to local nanoscale inhomogeneities such as Sr vacancies to stabilize a net switchable polarization in nanometer thick epitaxial, strain free SrTiO$_3$ film. The EDS study carried out by us also reveals Sr deficiencies at nanoscale. The nanopolar regions in our case should however be destabilized by the depolarizing field being isolated in the insulating matrix. When these nanopolar regions near the surface come in contact with metal electrodes, the external charges should be able to screen the depolarizing field thus leading to a switchable net electric polarization. Such a scenario should allow coherent rotation of the nanoscale dipoles under relatively low electric field, which explains the observed switching dynamics.
\section{Conclusion}
To summarize, ferroelectric relaxation is observed in STO single crystal which persists up to near room temperature. The switching polarization decreases with temperature while the coercive field and the relaxation time show anomalous temperature dependence. The associated switching dynamics is consistent with electric field induced coherent rotation of nano-polar domains near the electrode/$SrTiO_3$ interface.

\section{Acknowledgments}
VKS thanks CSIR, India for providing financial support.

\end{document}